\documentclass[12pt]{article}
\usepackage{amsfonts}
\usepackage[slantedGreek]{mathpazo}
\usepackage{graphicx,float,array}
\usepackage{amsmath,amssymb,amsthm,hyperref}
\usepackage[margin=1in]{geometry}
\usepackage{multirow}
\usepackage{hyperref}
\usepackage{algorithm, algorithmic}
\usepackage[utf8]{inputenc} 
\usepackage{caption}
\usepackage{marginnote}
\usepackage{xcolor}
\usepackage[round]{natbib}

\newcommand{\ALOOP}[1]{\ALC@it\algorithmicloop\ #1%
  \begin{ALC@loop}}
\newcommand{\ENDALOOP}{\end{ALC@loop}\ALC@it\algorithmicendloop}

\graphicspath{ {fig/} }

\begin{document}	
\title{Deep Learning Gaussian Processes For  Computer Models with Heteroskedastic and High-Dimensional Outputs}
\author{	
	\makebox[.4\linewidth]{Laura Schultz}\\
	\textit{ Department of Systems Engineering }\\
	\textit{  and Operations Research}\\
	\textit{George Mason University}\\
	\textit{email: lschult2@gmu.edu}
	\and 
	\makebox[.4\linewidth]{Vadim Sokolov\footnote{Corresponding author}}\\
	\textit{ Department of Systems Engineering }\\
	\textit{  and Operations Research}\\
	\textit{ George Mason University}\\	
	\textit{email: vsokolov@gmu.edu}
}
\date{First Draft: December, 2020\\This Draft: September 5, 2022}
\maketitle

\begin{abstract}
	Deep Learning Gaussian Processes (DL-GP) are proposed as a methodology for analyzing (approximating) computer models that produce heteroskedastic and high-dimensional output. Computer simulation models have many areas of applications, including  social-economic processes, agriculture, environmental, biology, engineering and physics problems. A deterministic transformation of inputs is performed by deep learning and predictions are calculated by traditional Gaussian Processes. We illustrate our methodology using a simulation of motorcycle accidents and simulations of an Ebola outbreak. Finally, we conclude with directions for future research. 
\end{abstract}
\section{Introduction}\label{Introduction}

Computer simulation experiments have gained popularity among engineering and science disciplines for the modeling and study of complex processes, such as manufacturing design, financial forecasting, environmental, and human system interactions. Supercomputers now allow for more realistic models, such as the popular Agent-Based Models (ABMs)~\cite{banks2021statistical,auld2016polaris,auld2012internet}, which simulate human behavior from a high-dimensional set of inputs and produce a large number of outputs. The complicating attribute of these simulators is their stochastic nature and heteroskedastic behavior, where noise levels depend upon the input variables~\cite{binois_practical_2018,schmidt_considering_2011,gelfand2004nonstationary}. Common types of analysis with stochastic simulators include sensitivity analysis, prediction, optimization and calibration. 

Traditional Monte Carlo-based approaches rely on repeated runs of each scenario for analysis~\cite{baker_analyzing_2020}; however, high-dimensional state-spaces are exponentially large and render most sampling methods ineffective. Another caveat is an inability to calculate derivatives of the simulator's input-output relationship and its stochastic nature makes formal statistical inference~\cite{more_benchmarking_2009,borgonovo_global_2022} inapplicable in these settings. Alternatively, the Bayesian surrogate approach~\cite{sacks_design_1989, kennedy_bayesian_2001}, which uses a statistical surrogate model to approximate the simulator~\cite{gramacy_adaptive_2009, snoek_input_2014, danielski_gaussian_2013, rasmussen_gaussian_2006, romero_navigating_2013}, has an analytical likelihood that can be used to infer behaviors in unevaluated regions. In this paper we propose a Bayesian framework to solve the prediction problem for simulators that are high dimensional and have multiple heteroskedastic outputs.

Our Bayesian approach places a Gaussian Process (GP) prior over the simulator's outputs and calculates the posterior using the results of initial model runs. The Bayesian approach allows for explicit quantification of uncertainty in unobserved input regions and the non-parametric GP model can learn complex input-output relations. However, the technique relies heavily on the informational contribution of each sample point and quickly becomes ineffective when faced with significant increases in dimensionality~\cite{shan_survey_2010,donoho_high_dimensional_2000}. Further, commonly fielded GP models assume that output variance is homogeneous. However, this assumption often proves unrealistic in practice and, therefore, the homogeneous models predict poorly~\cite{binois2018practical}. Unfortunately, the consideration of each input location to handle these heteroskedastic cases result in analytically intractable predictive density and marginal likelihoods~\cite{lazaro_gredilla_sparse_2011}. Furthermore, the smoothness assumption made by GP models hinders capturing rapid changes and discontinuities in the input-output relations. Popular attempts to overcome these issues include relying on the selection of kernel functions using prior knowledge about the target process~\cite{cortes_rational_2004}; splitting the input space into sub-regions so that inside each of those smaller subregions the target function is smooth enough and can be approximated with a GP model ~\cite{gramacy_bayesian_2008, gramacy_local_2015, chang_fast_2014}; and learning spatial basis functions~\cite{bayarri_framework_2007,wilson_fast_2014,higdon_space_2002}.

Another important feature of many practical computer models is that they have high-dimensional outputs. A naive approach to dealing with this is to place Gaussian priors to each of the outputs \cite{conti2010bayesian}. However, this approach ignores the correlation structure among the outputs, making learning less efficient~\cite{caruana_multitask_1997,bonilla_multi-task_2008} and can be computationally expensive when the number of outputs is large. Another approach~\cite{gattiker2006combining} is to assume the Kronecker structure in the simulation outputs, but this approach has limited applicability due to the constraints it adds to the form of the GP covariance function and assumption that data is iid. 

An alternative technique builds on the Linear Models of Coregionalization (LMC) approach originally used to model non-stationary and heteroskedastic spatio-temporal processes~\cite{mardia_spatial-temporal_1993, goulard1992linear, gelfand_nonstationary_2004}. A linear mixture of independent regression tasks are combined with coregionalization matrices to capture input-output correlations\cite{teh_semiparametric_2005,bonilla_multi-task_2008,osborne2009gaussian}. A primary advantage of this technique is the ability to use standard GPs, which assume stationary and isotropic variance, to produce a non-separable, non-stationary, and anisotropic estimation \cite{reich_class_2011}. There are several approaches to construct such a cross-covariance function for multiple output problems. For example, \cite{myers_co-kriging_1984} proposed multi-output functions which accounts for potential interdependence and use the LMC technique;  Convolutional Processes (CP) have been adapted by convolving univariate regression tasks with different smoothing kernel functions~\cite{higdon_space_2002, barry_blackbox_1996,alvarez2019non}; while, in the field of machine learning, Multi-task GPs construct a secondary covariance function~\cite{bonilla_multi-task_2008, alvarez_computationally_2011} between outputs. However, these approaches quickly grow unwieldy at high dimensions due to their additional correlation function in the order of $p(p+1)/2$ for $p$ outputs. In addition, their smoothness assumptions still hinder capturing rapid slope changes and discontinuities. For a recent discussion see \cite{genton2015cross}. 

One advantage of our approach is that we extend the coregionalization technique to handle functional, heteroskedastic computer model outputs. We add a pre-processing step that uses deep learning model to transform the training data set as previously proposed ~\cite{schultz_bayesian_2018,schultz2022bayesian,bhadra2021merging,polson2021deep,nareklishvili2022deep,wikle2019comparison,wikle2022statistical}. Our model uses deterministic transformation of the input $\theta$ defined by a deep learner $\psi = \phi(\theta)$ and then assumes that the heteroskedastic output $y$ is a weighted sum of the basis vectors $F(\psi)$ with $F(\psi)$ being the univariate Gaussian Processes. We define an estimation procedure that jointly estimates the parameters of the deep learning transformation, the basis vectors and hyperparameters of the univariate GPs. The goal of the deterministic transformation $\phi$ to find a vector representation of the input sets of the parameters so that dimensionality of this representation is lower compared to dimensionality of the output. Further, the basis vectors $F$ model correlation structure among elements of the output vector $y$. Further, we use an additional noise variable $e$ assigned to each of the $p$ outputs to account for any independent variations in the outputs. The model can now capture conditional relationships and any non-stationary and anisotropic behaviors without making the linear assumptions of other LMC methods or introducing new variable constraints~\cite{higdon_space_2002,reich_class_2011}. Further, this approach explicitly captures the uncertainty in the relations between the transformed inputs $\psi$ and the output $y$. By doing this, we improve our dimensionality reduction by introducing uncertainty to the latent features $\psi$, and further, we capture the uncertainty in the outputs $y$. By adding a deep learning pre-processing with Gaussian Process allows us to model discontinuities and steep slope changes. 

The remainder of the paper is organized as follows. In Section \ref{sec:meth}, we outline our modelling approach in detail. In Section \ref{sec:address}, we discuss various, limiting GP assumptions and illustrate how our model circumvents them. In Section \ref{sec:ebola}, we demonstrate our approach and apply our methodology to the epidemic model data produced by a stochastic, agent-based model (ABM) used in \cite{fadikar2018calibrating}. Finally, we provide a summary and potential opportunities for future work in Section \ref{sec:conclusion}.

\section{Background}

For context, we briefly review the GP regression method for a univariate output $y$ given observed input-output pairs $\mathcal{D} = \{\theta^{(i)},y(\theta^{(i)})\}_{i=1}^N$. The surrogate model seeks to approximate $y$ as follows,
\begin{equation}\label{eq:fx+ep+e}
y_i = F(\theta_i) + \epsilon(\theta) + e_i.
\end{equation}
The error $\epsilon(\theta)$ captures any parameter uncertainty, model inadequacy, and residual variability~\cite{kennedy_bayesian_2001}; and ${e}$ represents the observation error and residual variation inherent to the true process which generates $y$. 

Traditionally, Gaussian Process surrogates have been used for modeling $F$ by assigning a GP prior over a space of smooth functions, then updating this prior using the observed data $\mathcal{D}$. The resulting posterior distribution accounts for the uncertainty at unobserved points in the domain when performing interpolation and prediction; a GP is fully characterized by its mean function $m$ and covariance function, or kernel, $k$~\cite{gramacy_particle_2011}.
\label{key}
\begin{equation}
F(\theta)  \sim \mathcal{GP}\left(m(\theta),k(\theta)\right),
\end{equation}
where $k(\theta)_{ij} = k(\theta_i,\theta_j),~i\ne j$. 

It is common to model the prior mean by setting it to zero or via regression
\[
\mu(\theta) = h(\theta).
\]
The correlation equation $k(\theta)$ must result in a positive semi-definite matrix; in this paper, we use the Squared Exponential kernel, which is both stationary and isotropic:

\[ k_{SE}(\theta,\theta^\prime) = \exp{\left[-\frac{1}{2}{\left(\frac{\theta-{\theta}^{\prime}}{\lambda }\right)}^{2}\right]} \]
where $\lambda$ represents the lengthscale hyperparameter.

General references can be found in \cite{abrahamsen_review_1997} and \cite{duvenaud_automatic_2014}. 

Additionally, the inadequacy errors of $\epsilon(\theta)$ in Equation \ref{eq:fx+ep+e} can be incorporated via a specialized Linear ``nugget'' $ r(\theta) = \mathcal{N}(0,I\sigma_{\epsilon}^2)$ ~\cite{gramacy_cases_2012}:

\begin{equation}\label{eq:nugget}
\mathcal{L}(\theta) \sim \mathcal{GP}\left(m(\theta), K(\theta) = k(\theta,\theta^{\prime})+r(\theta)I\right)
\end{equation}
where $I$ represents the identity matrix.

Gaussian Process regression then infers the posterior distribution over functions on the observed data $\mathcal{D}$. The final density is a univariate Normal:

\begin{equation}\label{eq:posterior}
\left[y(\theta)  \mid  \mathcal{D}, \theta,\Omega\right] \sim \mathcal{N} \left(\mu, \Sigma\right),
\end{equation} 
where $\Omega$ are the parameters of the kernel function, and with the following summary statistics:
\begin{equation}\label{N_pred}
\begin{split}
\mu &= m(\theta) + k(\theta)(K_{\mathcal{D}})^{-1}(y-m_\mathcal{D})\\
\Sigma &= k(\theta,\theta) + r(\theta)-k(\theta)(K_{\mathcal{D}})^{-1}k(\theta)^T
\end{split}
\end{equation}
where $k(\theta)=\left(k(\theta,\theta^{(1)}),\ldots,k(\theta,\theta^{(N)})\right)^T$.

\section{Deep Learning Gaussian Processes (DL-GP)}\label{sec:meth}   
When approaching a multi-variable response problem, simply assigning a unique GP to each output may seem logical. However, the computational cost would be unmaintainable in large dimensions and the potential conditional dependencies among outputs lost. We propose incorporating a non-linear extension of the projection-based methods by using a deep learner model to construct a nonlinear mapping $\phi(\theta) \rightarrow \psi$ that transforms the input vector $\theta$ into a lower-dimensional vector of latent variables $\psi \in \mathbb{R}^q, q \ll p$ so that a basic GP can be used to model the relations between $\psi$ and $y$. A univariate prior $f(\cdot)$ is placed on each of these $q$ variables. The subset of GPs are then linearly combined to produce probabilistic estimates of the simulated outputs $y \in \mathbb{R}^p$ using a now-manageable number of GPs. Our overall surrogate model is as follows
\begin{align}
\psi & = \phi_{W}(\theta) \label{eq:nn} \\ 
F(\theta) & = b + \mathcal{W}f(\phi_{W}(\theta)) + e \label{eq:eta}\\
f_j(\psi_j) &\sim \mathcal{GP}\left(m_j(\psi_j),K_j(\psi_j, \psi_j \mid \Omega)\right),~j=1,\ldots,q \label{eq:gp} \\
e_i &\sim N\left(0, \sigma_{e_i}\right),~i=1,\ldots,p \label{eq:yerror} 
\end{align}

where $\mathcal{W} = (\mathcal{W}_1,\ldots,\mathcal{W}_q)$ are the basis vectors; GPs $f_j$ define weights over those basis vectors to represent the output $y$; $e$ represents the observational and residual errors of the true function; and $\phi$ is a deterministic non-linear map. For simplicity of notation, we will use $\psi$ instead of $\phi_{W}(\theta)$ moving forward. 

The simulated outputs then have a prior distribution of:

\begin{equation}
\begin{split}
F_i(\theta) &\sim \mathcal{GP}(M_i,S_i),~i=1,\ldots,p\\
M_i &= b_i + \sum_{j=1}^{q}\mathcal{W}_{ij}m_j(\psi_j)\\
S_i &= \sum_{j=1}^{q}\mathcal{W}_{ij}^2 K_j
\end{split}
\end{equation}

This prior is then integrated over the available data of $N$ observed input-output pairs $\mathcal{D} = (\psi^{(N)},y^{(N)})$ to compute the predictive posterior distribution for unobserved designs $\psi$:

\begin{equation}\label{eq:mvnpred_posterior}
F(\theta)  \mid  \mathcal{D}, \theta, \Omega \sim \mathcal{N} \left(\mu,\Sigma\right),
\end{equation} 
where $\Omega$ are the parameters of the kernel function, and with the following summary statistics:
\begin{equation}\label{mvn_pred}
\begin{split}
\mu &= M(\psi) + S(\psi)[S(\psi_N)+I\sigma_e^2]^{-1}(y_N-M(\psi_N))\\
\Sigma &= S(\psi,\psi) + e-S(\psi)[S(\psi_N)+I\sigma_e^2]^{-1}S(\psi)^T
\end{split}
\end{equation}
where $S(\psi)=\left(S(\psi,\psi^{(1)}),\ldots,S(\psi,\psi^{(N)})\right)^T$.

The diagram below visualizes our model.
\begin{figure}[H]
\centering
\includegraphics[width=\linewidth]{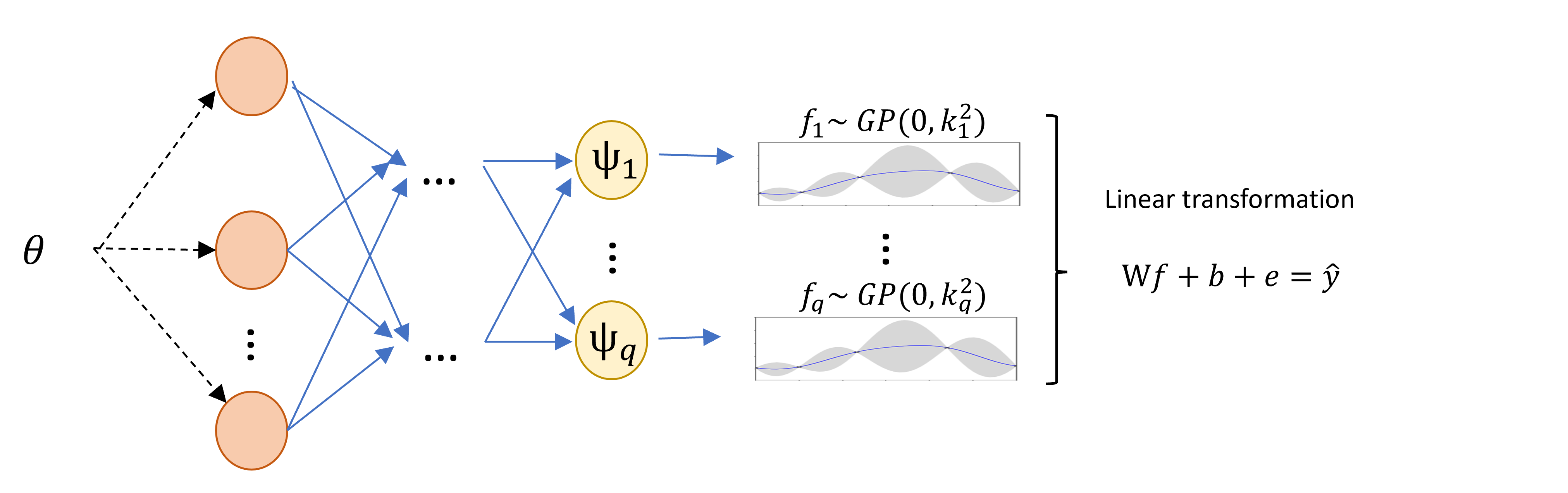}
\caption{Visualization of the proposed DL-GP model.}\label{fig:MO_GP}
\end{figure}

The conditional density for each output now depends on the linear weighted sums of a reduced number of manageable GPs, each having unique input sets. This setup allows for conditional dependencies to be captured in the posterior for the latent GPs and shares statistical strength across outputs. Further, the separate, Gaussian error term $e_i$ from Equation \ref{eq:yerror} allows the system's variance unique to the particular output to be captured and allow for conditional independence.

This mixture model of latent variables is the basis of our model and analogues to the Semi-Parametric Latent Variable Model (SPLM) and Factor Analysis Model~\cite{teh_semiparametric_2005}. However, we add the additional processing step of a nonlinear projection of the latent variables.

\subsection{Latent Variable Model}
The quality of surrogate's approximation depends on the quality of the training sample available. The goal is to capture the interaction behavior across the entire state-space; thereby making the placements and volume of samples critical. The volume of the input space grows exponentially as dimensionality increases. This `'curse of dimensionality'' leads to poor performance of non-parametric methods when inputs are high-dimensional due to the equal growth in sample volumes. Consequently, while capable of reproducing a wide range of behaviors and providing a highly flexible fit~\cite{gramacy_particle_2011}, GPs are known to fail in higher dimensions when resources are limited. Hence, we also augment the methodology with a deep learner to provide sufficient dimensionality reduction to combat sample restrictions.

Dimensional reduction is the key approach to solve the issue of high-dimensional outputs. An earlier approaches to dimensionality reduction included sparse models, which assume some local structures and that leads to a sparse covariance matrix. Another popular technique is variable selection, which choose a subset of variables providing the largest signal-to-noise ratio. However, both approaches are prone to overfitting and noise accumulation as dimensions rise~\cite{fan_challenges_2014}. A projection-based technique finds a condensed representation of the dataset in a lower dimensional subspace and have traditionally been implemented due to their computational and interpretable ease. For linear instances, one may use projection-based techniques such as Principal Component Analysis (PCA)~\cite{jolliffe_principal_2002}, Partial Least Squares (PLS)~\cite{abdi_partial_2003,polson2021deep}, or various single-index models ~\cite{adragni_sufficient_2009}. 

Deep Learning networks are a class of non-linear functions which construct a predictive mapping using hierarchical layers of latent variables. The singular or multivariate output can be a continuous, discrete, or mixed value set. Each layer $\ell$ applies, element-wise, a univariate activation function $\tau$ to an affine transformation:
\begin{equation}\label{basics}
\begin{split}
Z^{(1)} &=\tau^{(1)}\left(  W^{(0)}\Theta + b^{(0)}\right), \\
\cdots &\\
Z^{(\ell)} &=\tau^{(\ell)}\left(W^{(\ell -1)}Z^{(\ell -1)} + b^{(\ell - 1)}\right), \\
\hat{Y} & = W^{\ell}Z^{\ell}  + b^{(\ell)},
\end{split}
\end{equation}
where $W$ represents the weights placed on the layer's input set $Z$, and $b$ represents the offset value critical to recovering shifted multivariate functions.

Given the number of layers $\ell$, the multi-layered predictor is a composite map of cascading transformations

\begin{equation}\label{Eq:MLP}
Y = \left ( \tau_1^{W_1,b_1} \circ \ldots \circ \tau_\ell^{W_\ell,b_\ell} \right ) (\theta)
\end{equation}

The general approximation capabilities of deep learning networks is rooted in the premise first outlined by Kolmogorov~\cite{kurkova_kolmogorovs_1992}: any multivariate function $h(x)$ can be alternatively expressed as a finite composition of univariate functions and affine transformations\footnote{an affine transformation is a vector of a linear transformation plus an offset constant}. 

This is conceptually similar to the statistical \textit{ridge function} approximation paradigm \cite{chui_approximation_1992,le_mehaute_surface_1997}. Commonly referred to as Projection Pursuit Regression, the technique also asserts a multivariate function $h: \mathbb{R}^n \to \mathbb{R}$ can be approximately decomposed into a fixed set of $q$ directional hyperplanes known as \textit{ridge functions}:

\begin{equation}
h(\theta) = \sum_{i=1}^q g_i(a^i \cdot \theta),
\end{equation}
where $g_i: \mathbb{R} \to \mathbb{R}$ is a transformation function and $a^i \in \mathbb{R}^{n}\setminus\{0\},k<n$ is a directional matrix.

Deep learning networks can be alternatively viewed as a projection pursuit algorithm with nonlinear link functions. Each node effectively behaves as a dividing manifold and segregates the state-space collaboratively. In deep learning multi-layer networks, resultant stacking effect with each hidden layer is an exponential growth in the number of the system's dividing hyperplanes. As data progresses down the network, each layer applies a folding operator to predecessor's divided space. 

\begin{figure}[H]
    \centering
    \begin{tabular}{cc}
    \includegraphics[width=0.2\linewidth]{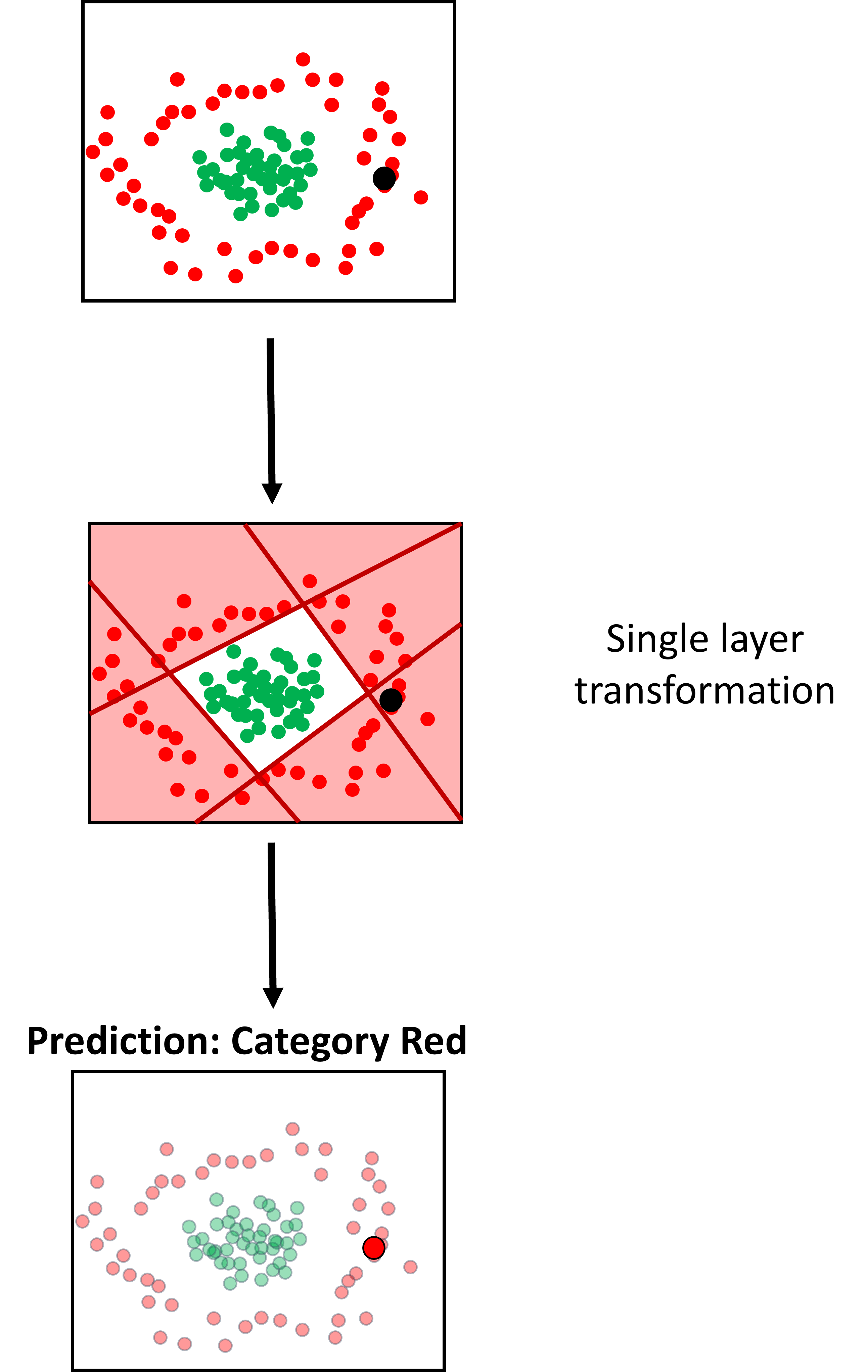} &
    \includegraphics[width=0.5\linewidth]{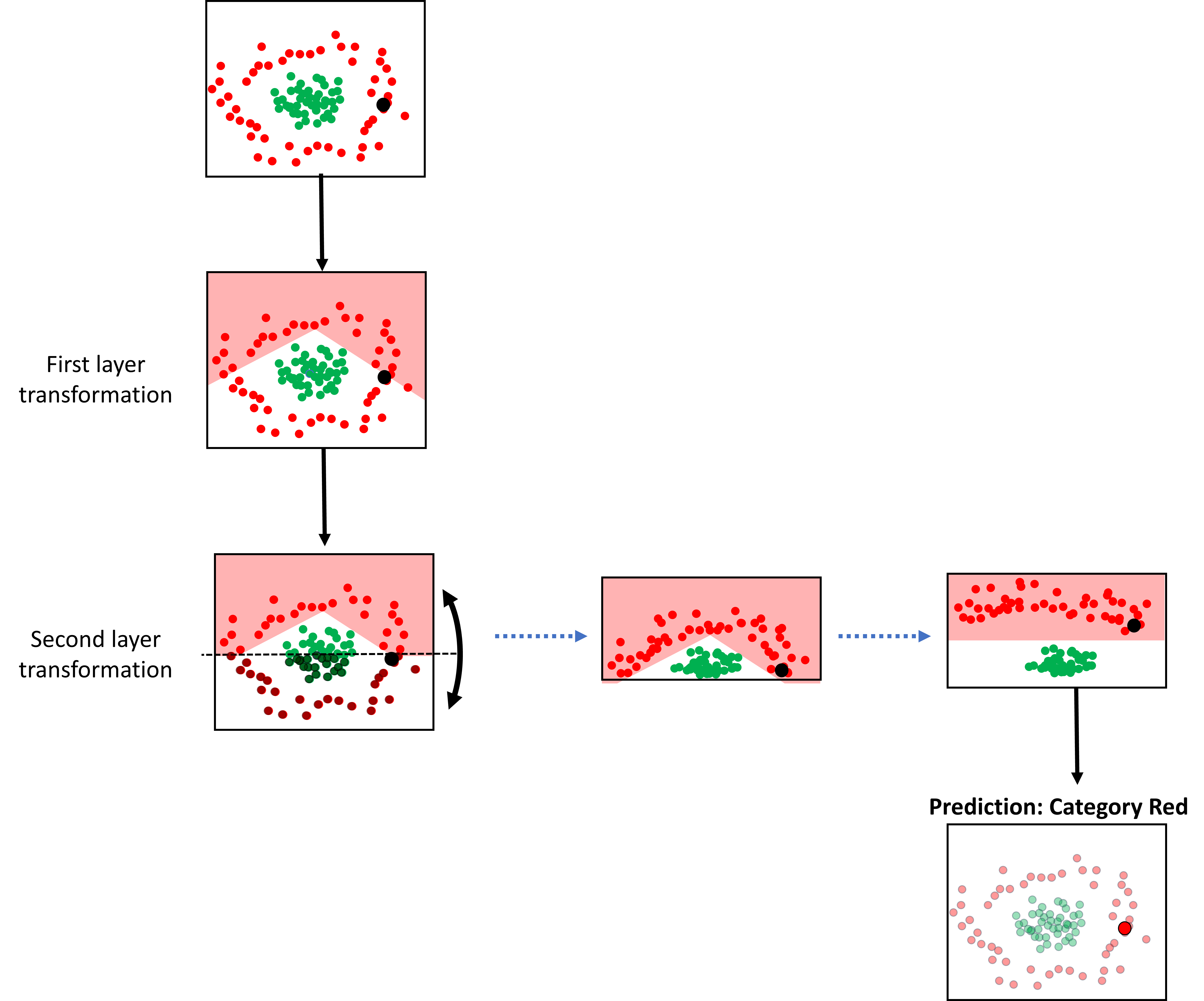} \\
    (a) Single-Layer Hyperplane Division & (b) Two-Layer Hyperplane Division
    \end{tabular}
    \caption{(a) A single layer network relies on hyperplanes (4 in this example) to partition the state space to correctly predict which group (red or green) the unknown dot (black) belongs to. (b) In a multi-layered network, each additional hidden layer allows the network to fold the developed hyperplane partitions of the previous layer independent of axis or orientation. The result is a need for less nodes in the previous layer (2 in this example) and ability to learn more complex patterns with less parameters.}
    \label{fig:hyperplanes}
\end{figure}

As shown in Figure \ref{fig:hyperplanes}, these recursive folds produce input-space regions independent of coordinate axes and, more pointedly, of different sizes~\cite{montufar_number_2014,pascanu_number_2013} in a significantly lower number of nodes~\cite{delalleau_shallow_2011}. 

Deep learning multi-layered networks designed for dimensionality reduction learning must contain at least one layer $Z^{(i)} \in \mathbb{R}^q$ with fewer nodes than the input or output layer of interest. This bottleneck layer forces the network to learn a compressed, or lower dimensional, set of latent features $psi$ describing the input-output relationship.

\begin{figure}
\centering
\begin{tabular}{cc}
\includegraphics[width=0.22\textwidth]{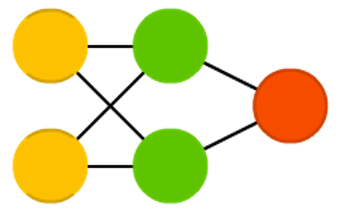} & \includegraphics[width=0.12\textwidth]{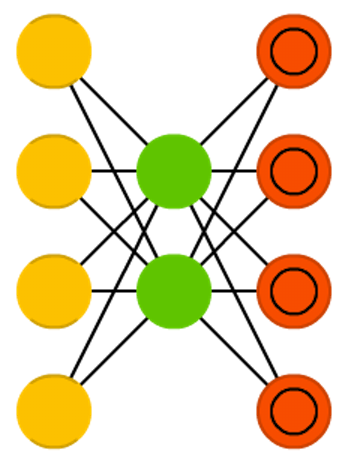}
\hspace*{16pt} \\
(a) General Architecture & (b) Bottleneck Example 
\end{tabular}
\caption{Each circle calculates a weighted sum of an input vector plus bias and applies a non-linear function to produce an output. Yellow and red colored neurons are input-output cells correspondingly. Source:	\url{http://www.asimovinstitute.org/neural-network-zoo}.}
\label{fig:arch}
\end{figure}

We use this representational form to extract a finite number of $q \ll p$ latent variables $\psi$ via the last hidden deep learner layer and incorporate our linearly mixed GP model by individually placing a univariate prior $f(\psi_j)$ on these unique feature mappings. The resulting GPs are then mixed using a final affine transformation, as documented in Equation \ref{eq:eta}.

Consequently, the error term for each single response variable $psi$ does not correspond to the original, principal subspace of the observed data but to the strength of the nonlinear projections in accounting for simulation discrepancy sources like parameter uncertainty, model inadequacy, and residual variability~\cite{kennedy_bayesian_2001}.

\subsection{Model Estimation}
Our composite model given by Equations (\ref{eq:nn}) - (\ref{eq:yerror}) is estimated using a maximum likelihood approach. The goal is to find the parameters $\{\Omega, W, \mathcal{W},b\}$ which best capture the given data $\mathcal{D}$, where $\Omega$ are the hyperparameters of the kernel functions of Gaussian processes $f_j~,j=1,\ldots,q$, $W$ are the weights of the deep learning function $\phi$ and $\mathcal{W}$ is the matrix the basis vectors that approximate the $y$-space.

Estimating these parameters represents a non-convex optimization problem. Deep learning model parameters are typically solved through a first-order method called \textit{gradient descent}, or \textit{backpropagation}. The gradient of a selected measure of error $\xi$, such as Maximum Likelihood Error (MSE) or Cross-Entropy Error, with respect to every weight and bias is found and used to update the original weights and biases:
\begin{align}
W^\ell & = W^\ell - \alpha \frac{\partial \xi}{\partial W^\ell}.
\end{align}

Our model's error function $\xi$ focuses on minimizing the log marginal likelihood.


\begin{equation}\label{eq:log_likelihood}
    L(y) \propto -\dfrac{1}{2}y^T\left(S^{-1}y - -\frac{1}{2} \log |S| - \frac{N}{2} \log(2\pi)\right\}
\end{equation}
where $F(\theta) \sim N(0,S)$.

While backpropogation can be applied to both the deep learning and mixed model weights and biases, the aggregation and resulting volatility in gradient methods do not provide satisfactory estimates for the GP hyperparameters. Particular to our paper's example, high time-correlated signals can produce severe vanishing gradient effects and cause convergence issues in gradient descent methods~\cite{gelbukh_nonlinear_2008}. Singly, GP implementations often approximate their unknown parameters through deterministic methods, such as Maximum Likelihood Estimate (MLE) or Maximum A Priori (MAP) estimates~\cite{gibbs_efficient_1997} or Markov Chain Monte Carlo (MCMC) sampling methods, such as Gibbs or Metropolis-Hastings. This paper excludes the GP-related hyperparameters from gradient descent training and uses a more generalized Monte Carlo method known as Slice Sampling~\cite{neal2003slice} at user-defined training intervals. This also allows our method to apply GPs which maintain non-Gaussian likelihoods.

The slice sampling algorithm~\cite{neal_bayesian_1996} seeks to generate samples from a target distribution with density $\pi(\theta)$ by introducing a uniform auxiliary variable $u \sim \mathcal{U}\left(0,\pi(\theta)\right)$.  
A horizontal line is drawn within $\pi(\theta)$ to define a horizontal `slice' $S = \{\theta:u < \pi(\theta)\}$ and an interval $I$ is formed by their union. A Markov chain is constructed by alternately updating $u$ and $\theta$ until the joint distribution $p(\theta,u)=p(u \mid \theta)p(\theta)$ becomes invariant. The auxiliary variable can then be discarded to provide only the sample of $\pi(\theta)$, the resulting marginal distribution. Note, since the target distribution $p(\theta)$ is unknown, a function $h(\theta)$ proportional to the distribution is substituted in the algorithm. 
\begin{algorithm}
\caption{Univariate Slice Sampling Algorithm}\label{algorithm_b}
\begin{algorithmic}[1]
\STATE Evaluate $h(\theta_0)$
\STATE Draw $u \sim \mathcal{U}(0,h(\theta_0))$ to define the horizontal ``slice'': $S = \{\theta:u < h(\theta)\}$
\STATE Create a horizontal interval $I= (L,R)$ enclosing $\theta_0$ 
\WHILE{true}
\STATE Draw $\theta_1 \sim \mathcal{U}(L,R)$
\STATE Evaluate $h(\theta_1)$
\IF {$h(\theta_1) > u$}
\STATE \textbf{break}
\ELSE
\STATE modify the interval $I$ according to a rejection procedure scheme
\ENDIF
\ENDWHILE
\STATE{\textbf{return} a new sample $\theta_1$}
\end{algorithmic}
\end{algorithm}

Ideally, the interval contains as much of the slice as feasible in order to allow the new point to 'jump' or differ significantly; however, if the interval is too large in comparison with the slice, the subsequent sampling step would become less efficient. Several schemes for finding the interval exist, with the two most commonly applied being the 'stepping out' and 'doubling procedures' to shorten or lengthen the scale based on the local features of the density distribution.

In a multivariate setting, the univariate slice sampling can be applied in a one-variable-at-a-time fashion but faces difficulty with highly correlated parameters. Neal proposed a multivariate version where a hypercube bounds the target slice $S$. While finding the appropriate interval is done through simultaneous adjustments in all dimensions, the step-size in each dimension can be unique, allowing local behaviors by variable to be reflected automatically. 

Other MCMC methods such as Gibbs sampling and Metropolis-Hastings algorithms would allow for sampling from multivariate distributions as well; however, Gibbs requires additional methods to sample from non-standard univariate distributions while Metropolis requires the user to find a precise 'proposal' distribution which can provide efficient sampling. In contrast, slice sampling avoids the need for precise tuning requirements since the algorithm's free variable, the size of the step when adjusting the sample interval, is chosen adaptively based on the local properties of the density function: if the step-size parameter is set too large, rejection procedures will exponentially tighten the interval towards the current sample while, if the value is too small, the procedures will adapt the reverse.

For this paper, we utilize the multivariate sampling method and define $h(\theta)$ as the log likelihood detailed in Equation \ref{eq:log_likelihood}.

\section{Non-Smooth and Heteroskedastic Data}\label{sec:address}

Although the additional nugget in Equation \ref{eq:nugget} allows us to capture stochasticity at a given point, GP models assume variable homogeneity across samples. Unfortunately, this assumption often proves unrealistic in practice but consideration of each input location to handle these heteroskedastic cases result in analytically intractable predictive density and marginal likelihoods ~\cite{lazaro2011variational}. In further complication, the corresponding assumption of Gaussian smoothness in the standard GP formulation hinders capturing rapid slope changes or discontinuities. 

Handling these behaviors is typically accomplished by using kernel functions that take advantage of the structure of the underlying process ~\cite{cortes_rational_2004}. For example, one can split the input space into sub-regions so that inside each of those smaller subregions the target function is smooth enough to be approximated with a GP model~\cite{gramacy_bayesian_2008,chang_fast_2014,kim_analyzing_2005}. Another approach is to learn spatial bias functions~\cite{bayarri_framework_2007, yang_bayesian_2015, wilson_fast_2014, wilson2013gaussian, higdon_space_2002}. Alternatively, there are adapted GP formulations crafted to specifically model heteroskedastic data ~\cite{gramacy_particle_2011, binois_practical_2018}; generally, the spatial relationship across points $e$ is modeled using a secondary GP~\cite{gelfand_nonstationary_2004} but at the cost of an additional correlation function in the order of $p(p+1)/2$ for $p$ outputs.

As noted by \cite{reich_class_2011}, the GP mixture weights of an LMC model would allow heteroskedastic behavior to be captured by automatically determining where specific feature mappings are most relevant. However, some apriori knowledge about the process is still required; for example, how many sub-regions must be captured. In our model, the deep learning multi-layered network can learn to turn certain mappings 'on' and 'off' independently of the mixture weights. Thus, we do not require any prior knowledge about the data structure. Further, the deep learning multi-layered network can transform steep slope changes into interpretable values for the GPs without introducing any linear constraints to the subspace projections~\cite{abdi_partial_2003,lynn_using_2000} or mixture weight restrictions like those found in \cite{reich_class_2011}, \cite{higdon_space_2002}, or \cite{teh_semiparametric_2005}.

\subsection{Motorcycle Data}

To illustrate, we use the heteroskedastic benchmark case detailed in \cite{binois_practical_2018}. In addition to being non-stationary, the problem possesses similar high time-correlated noise among the inputs to our paper's main problem. The example dataset samples the acceleration measurement $y$ from simulated motorcycle crashes at variable times $\theta$. Of the $N=133$ total samples recorded, $39$ use replicate inputs to capture the heteroskedastic nature of the simulation. 

\begin{figure}[h]
 	\centering
 	\includegraphics[width=.5\linewidth]{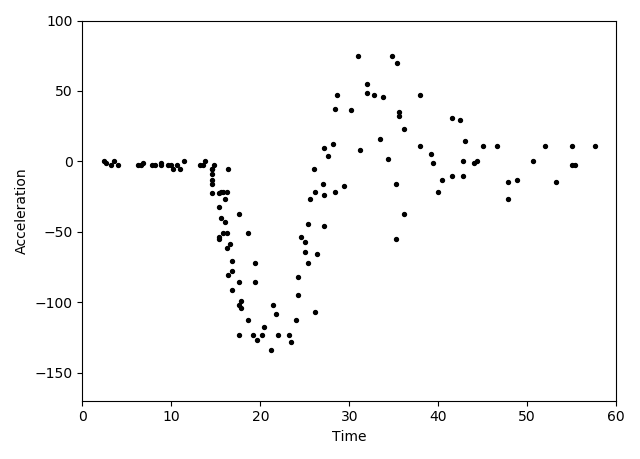}
 	\caption{Motorcycle example. Black dots represent the observed data points.}
 	\label{fig:motor_samples}
\end{figure}

A visual inspection of the data in Figure \ref{fig:motor_samples} shows there are three regions which exhibit distinct behavior among and across samples: the left portion has a constant variance and mean; the right portion has a larger but roughly constant variance with a gentle mean change; and the middle portion contains both steep and frequent variance changes along with a volatile mean. 

In step with the methods documented in \cite{binois_practical_2018}, we limit ourselves to a 2-dimensional GP structure and use the same Gaussian kernel to model the time-acceleration relationship. We then compare the resulting $90\%$ Confidence Interval (CI) against the results of two relevant models: a GP-only and DL-only framework. Our first benchmark is the heteroskedastic Gaussian Process (hetGP) formulation proposed in \cite{binois_practical_2018}. Their hetGP model produces a smoothed estimation of the nugget variable $r(\theta)$ from Equation \ref{eq:nugget} using the predictive mean of a secondary, regularizing GP. The construction of this secondary GP assumes a latent, log variation process. A joint log likelihood over both GPs allows for simultaneous optimization of all hyperparameters as well as the latent variables. For more explicit derivations, refer to \cite{binois_practical_2018}.  

For our second benchmark, we include a Deep Learning Quantile regression model, which draws from the Quantile Kriging approach established by \cite{plumlee2014building}. For a more recent review on Quantile Kriging and its applications, see \cite{baker_analyzing_2020}. The model predicts $5$ quantiles $(0.05, 0.20, 0.5,0.80, 0.95)$, or points under which a fraction of observations would fall below. For example, the $0.05$ quantile should over-predict $5\%$ of the time and contain $5\%$ of the observations in the area beneath it. The network is trained using a combination of the the Mean Squared Error (MSE) and quantile regression loss function:

\begin{equation}
 \frac{1}{N}  \sum_{i=0}^N \left(\left(y_i - \hat{y}_i\right)^2 + \max \left[q(\hat{y}_i - y_i), (q-1)(\hat{y}_i - y_i)\right]\right)
\end{equation}
where $q$ is the sought quantile. 

By defining the area between the $.95$ and $.05$ quantiles as the network's predicted $90\%$ confidence interval, the model provides a probabilistic answer despite its deterministic nature.

\begin{figure}[H]
\begin{tabular}{ccc}
\includegraphics[width=.3\linewidth]{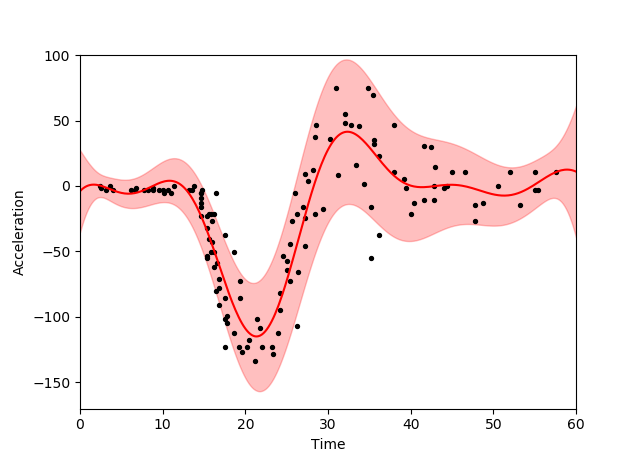} & 
\includegraphics[width=0.3\linewidth]{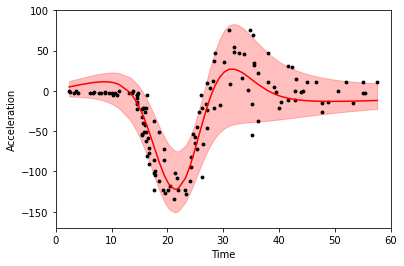} &
\includegraphics[width=0.3\linewidth]{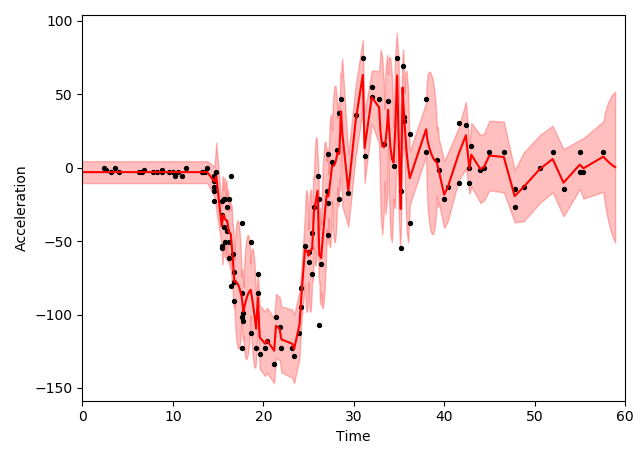}\\
(a) HetGP &(b) Deep Learning Quantile & (c) DL-GP \\
\end{tabular}
\caption{Motorcycle example. The model's fitted mean is displayed as a solid red line and 90\% confidence intervals are shown as red-filled regions. Black dots represent the observed data points.}\label{fig:motor_fit}
\end{figure}

The results of all three models are shown in Figure \ref{fig:motor_fit}. As hoped, every model, to varying degrees, successfully recognizes the heteroskedastic nature of the problem. Both the Quantile and hetGP methods provide a much smoother estimate over the statespace than our DL-GP but are unable to shadow the sharp, momentary drop in speed accompanying the function's peak like ours. In addition, our model is able to transition more rapidly to match the variance stabilization that happens on both ends of the timeline; our method is also able to reflect the nearly $90$ degree drop in acceleration about the $15$ second mark.

\begin{figure}[H]
\centering
\includegraphics[width=\linewidth]{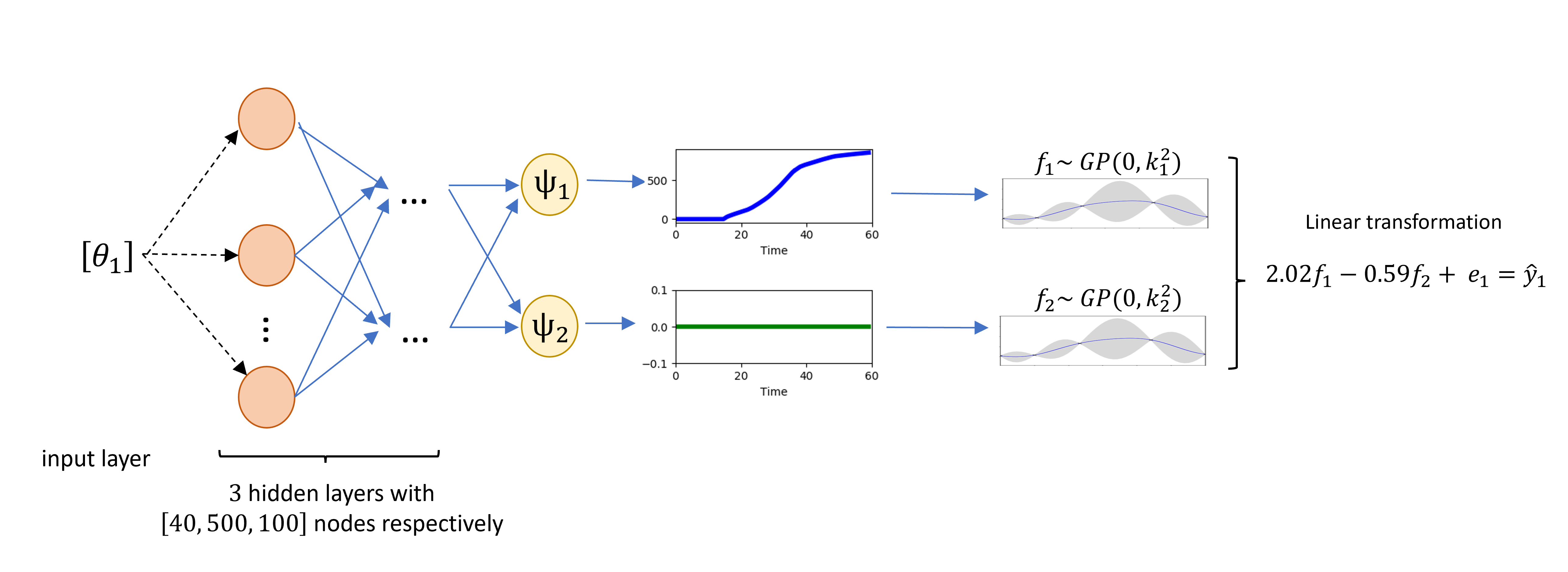}
\caption{The deep learning model produces two unique input sets to the model's two GPs. The first input set (blue) is variable across time while the second input set (green) remains constant. The weight (green rectangle) placed on the second GP's output would suggest the model seeks to eliminate a homoskedastic white noise present in the observations (black dots).}\label{fig:motor_weights}
\end{figure}

Figure \ref{fig:motor_weights} shows how the original subspace is transformed prior to the GP layer. The first GP is provided a variable input while the second GP is given a white-noise input. The lack of input changes in the second GP suggests that the additional dimension represents a source of homoskedastic noise found in the data. The negative weight assigned to its correspondent GP suggests that the deep learning feature serves as a counter to a systemic noise source like measurement error.

For quantifiable comparison, we follow the same experimental setup of $300$ independent splits generated by randomly subdividing the full dataset into $90\%$-$10\%$ training and testing sets. Our methodology's performances across partitions are used to generate the Normalized Mean Squared Error (NMSE) and Negative Log-Probability Density (NLPD) metrics detailed in the original \cite{lazaro2011variational} study:

\begin{equation}
\begin{split}
NMSE &= \frac{\sum_{j=1}^{n_\ast}\left(y_{\ast j} - \hat{y}_{\ast j}\right)^2}{\sum_{j=1}^{n_\ast}\left(y_{\ast j} - \bar{y}\right)^2}, \\
NLPD &= -\frac{1}{n_\ast} \sum_{j=1}^{n_\ast } \log \mathit{p}\left(y_{\ast j} \mid \mathcal{D}\right),
\end{split}
\end{equation}

where $y_{\ast j}$ is the $j$-th observation within the test set; $\hat{y}_{\ast j}$  is the mean of the posterior for that observation; $n_\ast$ is the number of test observations; and $\bar{y}$ is the mean of the training observations.

\begin{table}[H]
	\centering
	\begin{tabular}{c| c| c| c| c| c| c}
		& DL-GP & Q-DL & WHGP&  GP &MAPHGP & VHGP\\
		\hline
		NMSE & $0.20 \pm 0.07$ & $0.31 \pm 0.21$ & $0.28 \pm 0.21$ & $0.26 \pm 0.18$ & $0.26 \pm 0.17$ & $0.26 \pm 0.17$ \\
		\hline
		NLPD &$0.68 \pm 0.18 $ & N/A & $4.26 \pm 0.31$ &$4.59 \pm 0.22$ &$4.32 \pm 0.60 $&$4.32 \pm 0.30$ \\
	\end{tabular}
	\caption{Average Normalized Mean Squared Error (NMSE) and Negative Log-Probability Density (NLPD) $\pm 1$ standard deviation for $300$ random partitions. Lower values are desired.}
	\label{fig:motor_tab}
\end{table}

Table \ref{fig:motor_tab} summarizes the results for our proposed model (labeled DL-GP) and the quantile deep learner (labeled Q-DL) in comparison with those recorded in \cite{binois_practical_2018}: their smoothed heteroskedastic GP (labeled WHGP); a standard, homoskedastic GP (labeled  GP); a Maximum A Posteriori heteroskedastic GP (labeled MAPHGP) from \cite{kersting_most_2007}; and a variational model (labeled VHGP) from \cite{lazaro2011variational}. 

Note that our two models are comparable to these alternatives.

\section{Application: Predicting Ebola Epidemic ABM}\label{sec:ebola}

For illustration, we use the multi-output agent-based epidemic model problem documented in \cite{fadikar2018calibrating}. We predict the $56$-week, simulated outputs for three holdout scenarios.

\subsection{Data Set}
After the 2014-2015 West Africa Ebola outbreak, the Research and Policy for Infectious Disease Dynamics (RAPIDD) program at the National Institutes of Health (NIH) convened a workshop to compile and explore the various forecasting approaches used to help manage the outbreak. At its conclusion, a disease forecasting challenge was launched to provide 4 synthetic population datasets and scenarios as a baseline for cross-assessment. A stochastic, agent-based model~\cite{ajelli_rapidd_2018} first generated each population using varying degrees of data accuracy, availability, and intervention measures; individuals were then assigned activities based on demographic and survey data to model realistic disease propagation. Transmission by an infected individual is determined probabilistically based on the duration of contact with a susceptible individual and $d=5$ static inputs $\Theta = {\theta_1,cdots,\theta_5}$ to the model.

\begin{figure}[H]
	\centering
		\begin{tabular}{c|c|c} 
			Parameter & Description & Range \\ \hline
			$\theta_1$ & probability of disease transmission & $[3\times10^{-5}, 8\times10^{-5}]$ \\ 
			$\theta_2$ & initial number of infected individuals & $[1, 20]$ \\
			$\theta_3$ & delay in hospital intervention  & $[2,10]$ \\ 
			$\theta_4$ & efficacy of hospital intervention & $[0.1,0.8]$ \\  
			$\theta_5$ & intervention reduction of travel & $[3\times10^{-5}, 8\times10^{-5}]$ \\ 
		\end{tabular}
\caption{$5$ static inputs used for defining disease propagation for the Ebola ABM}
\end{figure}

A single run outputs a cumulative count of infected individuals over a $56$-week period. For more details on the model and challenge, see \cite{viboud_rapidd_2018}.

\begin{figure}[H]
	\centering
	\includegraphics[width=.7\linewidth]{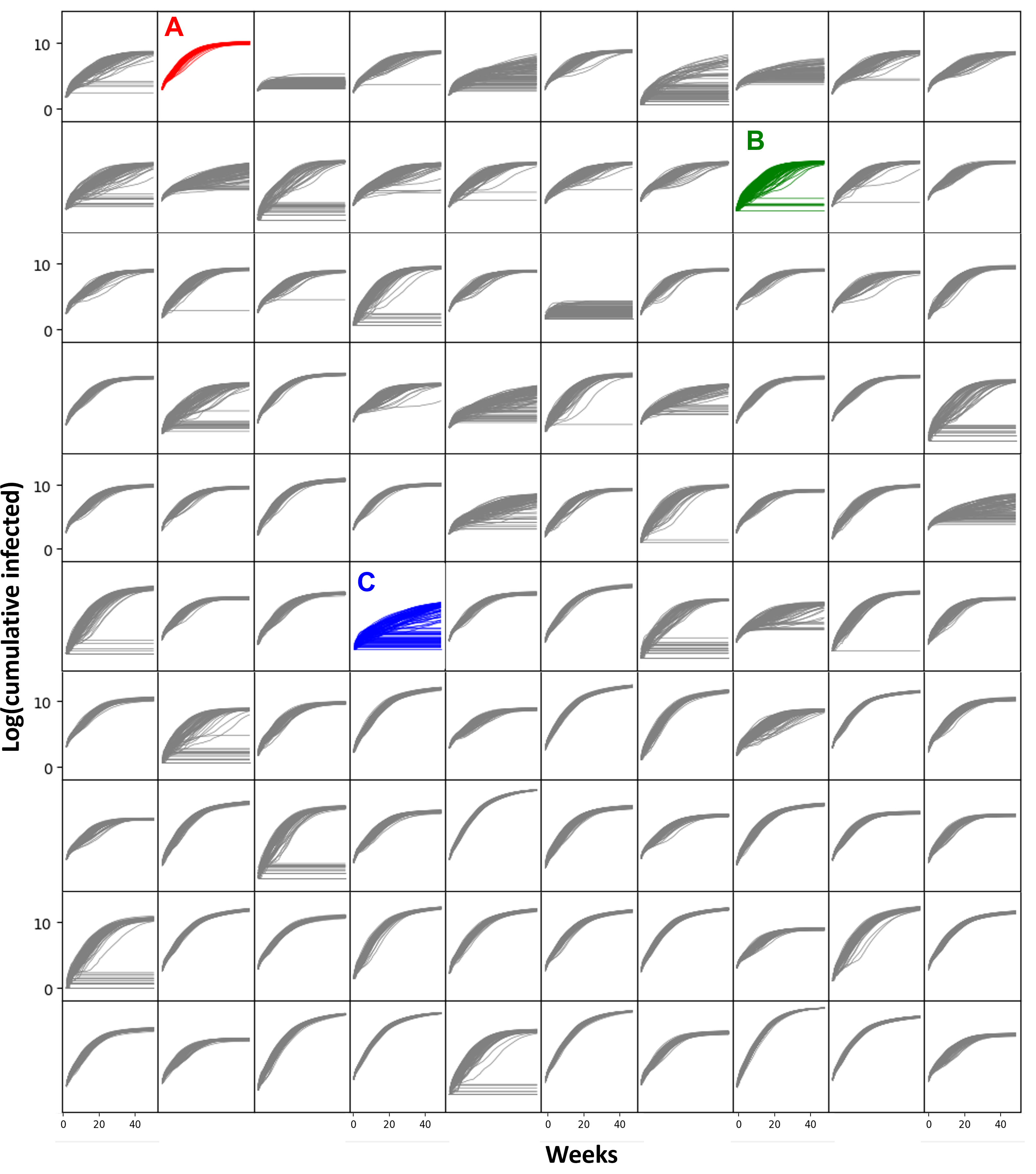}
	\caption{For each scenario, the $100$ simulated trajectories for the cumulative number of disease incidences across $56$ weeks are shown as grey lines. The three holdout scenarios (A, B, and C) are highlighted in red, green, and blue respectively.
	}\label{fig:all_samples}
\end{figure}

To maintain comparison, we use the same data set from \cite{fadikar2018calibrating}, which consists of a collection of $m=100$ scenarios generated through a space-filling, symmetric Latin hypercube design. For each scenario, $100$ replicates were run for a total of $N=10,000$ simulated epidemic trajectories. A single run for each parameter set produced a $56$-dimensional output, capturing the cumulative weekly number of infected individuals. The log results for every setting combination is captured in Figure \ref{fig:all_samples}. For testing purposes, we exclude the same $3$ unique parameter settings, which we will refer to as A, B, and C, and their respective $n=100$ simulated outputs from the training set.

Notably, different parameter settings produced strongly divergent behaviors in their replicates. Some followed a mean trajectory while others produced strong bimodal behavior, significant heteroskedasticity, or simply remained flat-lined.
\cite{fadikar2018calibrating} reasons that only a subset of replicates within each parameter settings may produce similar initial behaviors and that adding an additional variable $\alpha$ to index these replicates would produce more accurate predictions. Modifying the Quantile Kriging approach, the $100$ replicates of each parameter setting are replaced with $n_\alpha = 5$ quantile-based trajectories. The quantiles are then indexed within each parameter set by the addition of a sixth latent variable $\alpha\in [0,1]$:
\begin{equation}
\Theta = \left[\theta_1, \theta_2, \theta_3, \theta_4, \theta_5, \alpha \right].
\end{equation}

\begin{figure}[H]
 \centering
 \includegraphics[width=.5\linewidth]{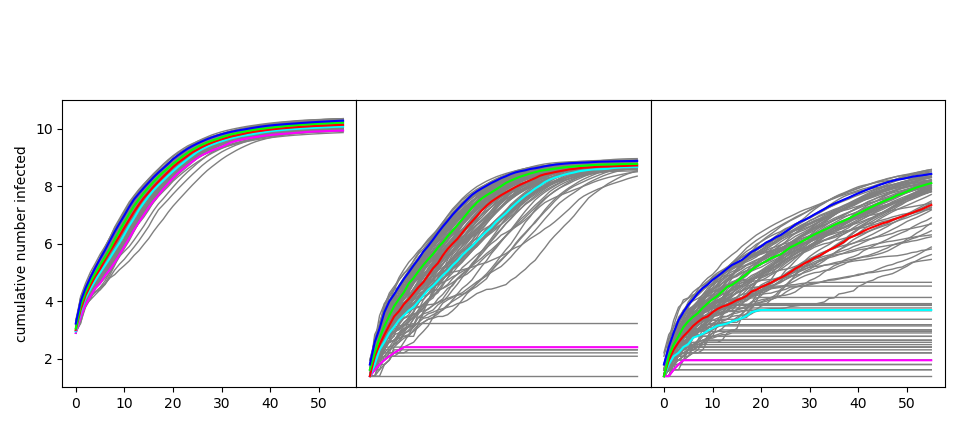}
 \caption{For each holdout scenario, labeled A, B, and C, the $100$ simulated trajectories of the cumulative number of disease incidences over $56$ weeks are shown as grey lines. The $5$ colored lines represent the $[0.05, 0.275, 0.5, 0.725, 0.95]$ quantiles, which are now indexed by the additional sixth parameter $\alpha$.}
 \label{fig:holdout_q}
\end{figure}

Figure \ref{fig:holdout_q} shows the $5$ calculated quantiles of these three hold-out scenarios, labeled A, B, and C and highlighted in red, green, and blue respectively, that we wish to predict using our model.

\begin{figure}[H]
 \centering
 \includegraphics[width=\linewidth]{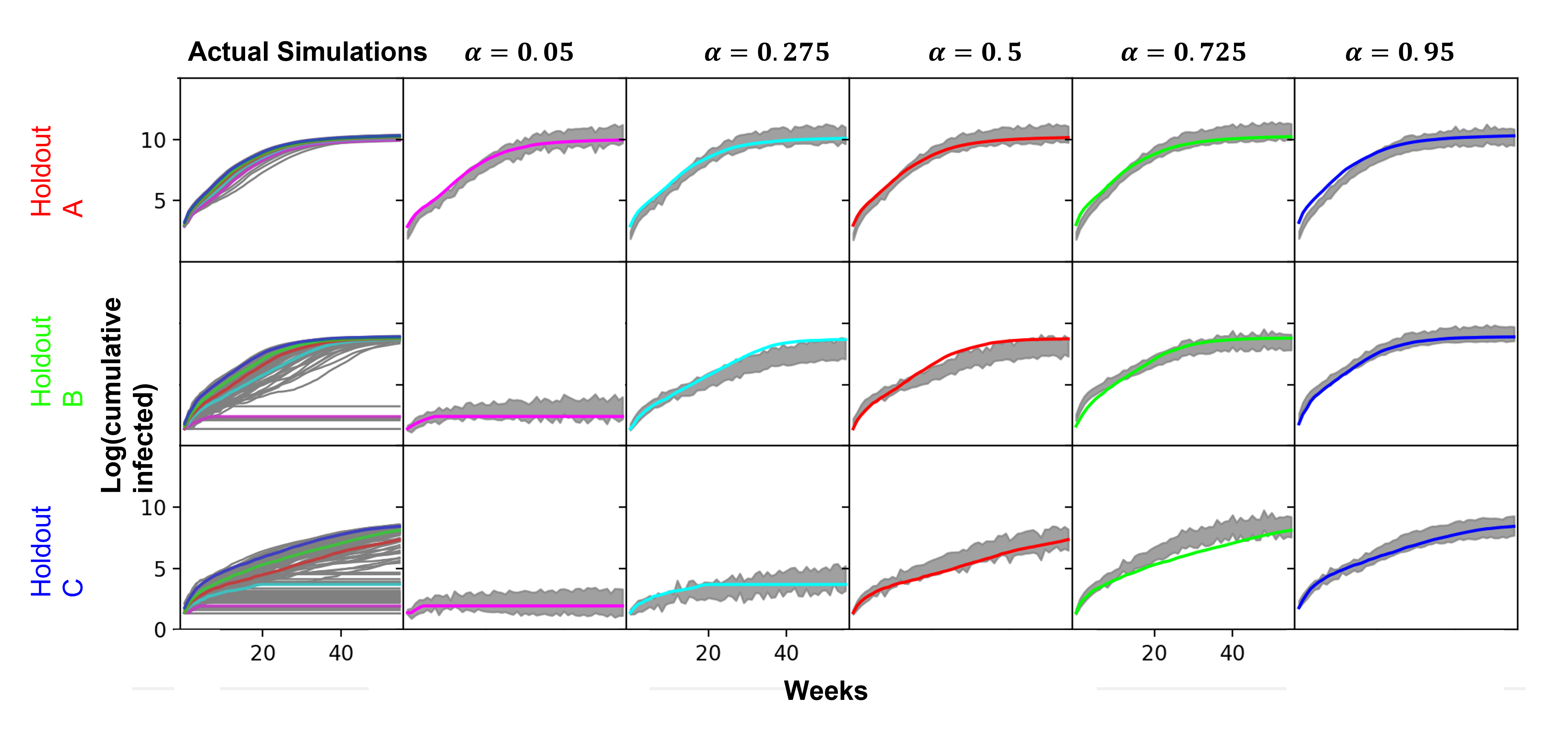}
 \caption{The first column shows the $100$ actual simulations and their empirical quantiles for each of the three holdout scenarios. The next $5$ columns show the predictive posterior's $90\%$ confidence intervals for each empirical quantile, denoted by different colors.}
 \label{fig:ebola_res}
\end{figure}

Figure \ref{fig:ebola_res} shows the results of our experiment. Each row of the figure corresponds to one of the holdout scenarios (A, B, and C); the first column shows the $100$ replicates generated by the ABM, along with the estimated quantiles; the remaining $5$ columns compare the $90\%$ confidence intervals our model predicts for each of the $5$ quantile settings estimated from the ABM replicates.

Our model's contribution to \cite{fadikar_calibrating_2017} successful approach is an improvement in capturing the saturation points of the various scenarios' disease trajectories with tighter confidence intervals, particularly during the first $20$ weeks when the the disease propagation is steepest, while still encompassing or closely aligning to each holdout quantile. Further, our model proves more adept at capturing the more linear growth behaviors among several cases, though we acknowledge that the linearity in the $q=.725$ quantile for scenario C is not well captured. 

Overall, our model's predictions among the lower quantiles provide more noticeable improvements, including the correction of several of \cite{fadikar_calibrating_2017} cases, to maintain the quantile across the entire timeline's 90\% CI. 
In contrast, some of our results underestimated the initial case loads when \cite{fadikar_calibrating_2017}'s model did not; however, most of these cases incorporated the difference within the following weeks without missing the quantile's inflection points.

\section{Discussion}\label{sec:conclusion}

Deep Learning Guassian Process (DL-GP) models provide a flexible class of high-dimensional input-output models that can achieve smoothing, prediction, and calibration in a variety of contexts. Each GP is built over the significant factors found by a non-linear projection of the simulation's inputs via a deep learning multi-layered network. This model allows us to leverage conditional relationships to help predict the outputs for high-dimensional simulators with multiple outputs without the constraints found in other methods. Furthermore, the non-linear transformation technique eliminates the unrealistic requirement of smooth, homoskedastic behavior without introducing linear constraints to the subspace projections or mixture weight restrictions.
There are many directions for future research. We wish to next incorporate this new model into a calibration framework to see how it performs against our previously developed high-dimensional model, which collapsed the outputs to a single predictive value before optimizing. Additional research of interest include incorporating a parallelized or mini-batch slicing algorithm that would improve computational timing and align with the paradigm of deep learning networks training.

\bibliography{IEEEfull_Bibtex}

\begin{thebibliography}{78}
\providecommand{\natexlab}[1]{#1}
\providecommand{\url}[1]{\texttt{#1}}
\expandafter\ifx\csname urlstyle\endcsname\relax
  \providecommand{\doi}[1]{doi: #1}\else
  \providecommand{\doi}{doi: \begingroup \urlstyle{rm}\Url}\fi

\bibitem[Abdi(2003)]{abdi_partial_2003}
Hervé Abdi.
\newblock Partial least square regression ({PLS} regression).
\newblock \emph{Encyclopedia for research methods for the social sciences},
  6\penalty0 (4):\penalty0 792--795, 2003.

\bibitem[Abrahamsen(1997)]{abrahamsen_review_1997}
Petter Abrahamsen.
\newblock Gaussian random fields and correlation functions.
\newblock Technical report, Technical report, Norwegian Computing Center, Oslo,
  Norway, 1997.

\bibitem[Adragni and Cook(2009)]{adragni_sufficient_2009}
Kofi~P. Adragni and R.~Dennis Cook.
\newblock Sufficient dimension reduction and prediction in regression.
\newblock \emph{Philosophical Transactions of the Royal Society A:
  Mathematical, Physical and Engineering Sciences}, 367\penalty0
  (1906):\penalty0 4385--4405, November 2009.

\bibitem[Ajelli et~al.(2018)Ajelli, Zhang, Sun, Merler, Fumanelli, Chowell,
  Simonsen, Viboud, and Vespignani]{ajelli_rapidd_2018}
Marco Ajelli, Qian Zhang, Kaiyuan Sun, Stefano Merler, Laura Fumanelli, Gerardo
  Chowell, Lone Simonsen, Cecile Viboud, and Alessandro Vespignani.
\newblock The {RAPIDD} {Ebola} forecasting challenge: {Model} description and
  synthetic data generation.
\newblock \emph{Epidemics}, 22:\penalty0 3--12, 2018.

\bibitem[{\'A}lvarez et~al.(2019){\'A}lvarez, Ward, and
  Guarnizo]{alvarez2019non}
Mauricio~A {\'A}lvarez, Wil Ward, and Cristian Guarnizo.
\newblock Non-linear process convolutions for multi-output gaussian processes.
\newblock In \emph{The 22nd International Conference on Artificial Intelligence
  and Statistics}, pages 1969--1977. PMLR, 2019.

\bibitem[Arasaratnam and Haykin(2008)]{gelbukh_nonlinear_2008}
Ienkaran Arasaratnam and Simon Haykin.
\newblock Nonlinear {Bayesian} {Filters} for {Training} {Recurrent} {Neural}
  {Networks}.
\newblock In Alexander Gelbukh and Eduardo~F. Morales, editors, \emph{{MICAI}
  2008: {Advances} in {Artificial} {Intelligence}}, volume 5317, pages 12--33.
  Springer Berlin Heidelberg, Berlin, Heidelberg, 2008.

\bibitem[Auld et~al.(2012)Auld, Sokolov, Fontes, and
  Bautista]{auld2012internet}
Joshua Auld, Vadim Sokolov, Angela Fontes, and Rene Bautista.
\newblock Internet-based stated response survey for no-notice emergency
  evacuations.
\newblock \emph{Transportation Letters}, 4\penalty0 (1):\penalty0 41--53, 2012.

\bibitem[Auld et~al.(2016)Auld, Hope, Ley, Sokolov, Xu, and
  Zhang]{auld2016polaris}
Joshua Auld, Michael Hope, Hubert Ley, Vadim Sokolov, Bo~Xu, and Kuilin Zhang.
\newblock Polaris: Agent-based modeling framework development and
  implementation for integrated travel demand and network and operations
  simulations.
\newblock \emph{Transportation Research Part C: Emerging Technologies},
  64:\penalty0 101--116, 2016.

\bibitem[Baker et~al.(2020)Baker, Barbillon, Fadikar, Gramacy, Herbei, Higdon,
  Huang, Johnson, Ma, Mondal, Pires, Sacks, and Sokolov]{baker_analyzing_2020}
Evan Baker, Pierre Barbillon, Arindam Fadikar, Robert~B. Gramacy, Radu Herbei,
  David Higdon, Jiangeng Huang, Leah~R. Johnson, Pulong Ma, Anirban Mondal,
  Bianica Pires, Jerome Sacks, and Vadim Sokolov.
\newblock Analyzing {Stochastic} {Computer} {Models}: {A} {Review} with
  {Opportunities}, September 2020.

\bibitem[Banks and Hooten(2021)]{banks2021statistical}
David~L Banks and Mevin~B Hooten.
\newblock Statistical challenges in agent-based modeling.
\newblock \emph{The American Statistician}, 75\penalty0 (3):\penalty0 235--242,
  2021.

\bibitem[Barry and {Jay M. Ver Hoef}(1996)]{barry_blackbox_1996}
Ronald~Paul Barry and {Jay M. Ver Hoef}.
\newblock Blackbox {Kriging}: {Spatial} {Prediction} without {Specifying}
  {Variogram} {Models}.
\newblock \emph{Journal of Agricultural, Biological, and Environmental
  Statistics}, 1\penalty0 (3):\penalty0 297--322, 1996.

\bibitem[Bayarri et~al.(2007)Bayarri, Berger, Paulo, Sacks, Cafeo, Cavendish,
  Lin, and Tu]{bayarri_framework_2007}
Maria~J Bayarri, James~O Berger, Rui Paulo, Jerry Sacks, John~A Cafeo, James
  Cavendish, Chin-Hsu Lin, and Jian Tu.
\newblock A {Framework} for {Validation} of {Computer} {Models}.
\newblock \emph{Technometrics}, 49\penalty0 (2):\penalty0 138--154, May 2007.

\bibitem[Bhadra et~al.(2021)Bhadra, Datta, Polson, Sokolov, and
  Xu]{bhadra2021merging}
Anindya Bhadra, Jyotishka Datta, Nick Polson, Vadim Sokolov, and Jianeng Xu.
\newblock Merging two cultures: Deep and statistical learning.
\newblock \emph{arXiv preprint arXiv:2110.11561}, 2021.

\bibitem[Binois et~al.(2018{\natexlab{a}})Binois, Gramacy, and
  Ludkovski]{binois_practical_2018}
Mickael Binois, Robert~B. Gramacy, and Michael Ludkovski.
\newblock Practical heteroskedastic {Gaussian} process modeling for large
  simulation experiments.
\newblock \emph{Journal of Computational and Graphical Statistics}, 27\penalty0
  (4):\penalty0 808--821, October 2018{\natexlab{a}}.

\bibitem[Binois et~al.(2018{\natexlab{b}})Binois, Gramacy, and
  Ludkovski]{binois2018practical}
Mickael Binois, Robert~B Gramacy, and Mike Ludkovski.
\newblock Practical heteroscedastic gaussian process modeling for large
  simulation experiments.
\newblock \emph{Journal of Computational and Graphical Statistics}, 27\penalty0
  (4):\penalty0 808--821, 2018{\natexlab{b}}.

\bibitem[Bonilla et~al.(2008)Bonilla, Chai, and
  Williams]{bonilla_multi-task_2008}
Edwin~V Bonilla, Kian Ming~A Chai, and Christopher K~I Williams.
\newblock Multi-task {Gaussian} {Process} {Prediction}.
\newblock \emph{Advances in neural information processing systems}, page~8,
  2008.

\bibitem[Borgonovo et~al.(2022)Borgonovo, Li, Barr, Plischke, and
  Rabitz]{borgonovo_global_2022}
Emanuele Borgonovo, Genyuan Li, John Barr, Elmar Plischke, and Herschel Rabitz.
\newblock Global {Sensitivity} {Analysis} with {Mixtures}: {A} {Generalized}
  {Functional} {ANOVA} {Approach}.
\newblock \emph{Risk Analysis}, 42\penalty0 (2):\penalty0 304--333, 2022.

\bibitem[Caruana(1997)]{caruana_multitask_1997}
Rich Caruana.
\newblock Multitask {Learning}.
\newblock \emph{Machine Learning}, 28\penalty0 (1):\penalty0 41--75, July 1997.

\bibitem[Chang et~al.(2014)Chang, Haran, Olson, and Keller]{chang_fast_2014}
Won Chang, Murali Haran, Roman Olson, and Klaus Keller.
\newblock Fast dimension-reduced climate model calibration and the effect of
  data aggregation.
\newblock \emph{The Annals of Applied Statistics}, 8\penalty0 (2):\penalty0
  649--673, June 2014.

\bibitem[Chui and Li(1992)]{chui_approximation_1992}
Charles~K Chui and Xin Li.
\newblock Approximation by ridge functions and neural networks with one hidden
  layer.
\newblock \emph{Journal of Approximation Theory}, 70\penalty0 (2):\penalty0
  131--141, August 1992.

\bibitem[Conti and O'Hagan(2010)]{conti2010bayesian}
Stefano Conti and Anthony O'Hagan.
\newblock Bayesian emulation of complex multi-output and dynamic computer
  models.
\newblock \emph{Journal of statistical planning and inference}, 140\penalty0
  (3):\penalty0 640--651, 2010.

\bibitem[Cortes et~al.(2004)Cortes, Haffner, and Mohri]{cortes_rational_2004}
Corinna Cortes, Patrick Haffner, and Mehryar Mohri.
\newblock Rational kernels: {Theory} and algorithms.
\newblock \emph{Journal of Machine Learning Research}, 5\penalty0
  (Aug):\penalty0 1035--1062, 2004.

\bibitem[Danielski et~al.(2013)Danielski, Kacprzak, Tinetti, and
  Jagoda]{danielski_gaussian_2013}
C.~Danielski, T.~Kacprzak, G.~Tinetti, and P.~Jagoda.
\newblock Gaussian {Process} for star and planet characterisation.
\newblock \emph{arXiv:1304.6673 [astro-ph]}, April 2013.

\bibitem[Delalleau and Bengio(2011)]{delalleau_shallow_2011}
Olivier Delalleau and Yoshua Bengio.
\newblock Shallow vs. {Deep} {Sum}-{Product} {Networks}.
\newblock In J.~Shawe-Taylor, R.~S. Zemel, P.~L. Bartlett, F.~Pereira, and
  K.~Q. Weinberger, editors, \emph{Advances in {Neural} {Information}
  {Processing} {Systems} 24}, pages 666--674. Curran Associates, Inc., 2011.

\bibitem[Donoho(2000)]{donoho_high_dimensional_2000}
David~L. Donoho.
\newblock High-dimensional data analysis: {The} curses and blessings of
  dimensionality.
\newblock In \emph{Ams {Conference} on {Math} {Challenges} of the 21st
  {Century}}, 2000.

\bibitem[Duvenaud(2014)]{duvenaud_automatic_2014}
David Duvenaud.
\newblock \emph{Automatic model construction with {Gaussian} processes}.
\newblock PhD thesis, University of Cambridge, 2014.

\bibitem[Fadikar et~al.(2017)Fadikar, Higdon, Chen, Lewis, Venkatramanan, and
  Marathe]{fadikar_calibrating_2017}
Arindam Fadikar, Dave Higdon, Jiangzhuo Chen, Brian Lewis, Srini Venkatramanan,
  and Madhav Marathe.
\newblock Calibrating a {Stochastic} {Agent} {Based} {Model} {Using}
  {Quantile}-based {Emulation}.
\newblock \emph{arXiv:1712.00546 [stat]}, December 2017.

\bibitem[Fadikar et~al.(2018)Fadikar, Higdon, Chen, Lewis, Venkatramanan, and
  Marathe]{fadikar2018calibrating}
Arindam Fadikar, Dave Higdon, Jiangzhuo Chen, Bryan Lewis, Srinivasan
  Venkatramanan, and Madhav Marathe.
\newblock Calibrating a stochastic, agent-based model using quantile-based
  emulation.
\newblock \emph{SIAM/ASA Journal on Uncertainty Quantification}, 6\penalty0
  (4):\penalty0 1685--1706, 2018.

\bibitem[Fan et~al.(2014)Fan, Han, and Liu]{fan_challenges_2014}
Jianqing Fan, Fang Han, and Han Liu.
\newblock Challenges of {Big} {Data} {Analysis}.
\newblock \emph{National science review}, 1\penalty0 (2):\penalty0 293--314,
  June 2014.

\bibitem[Gattiker et~al.(2006)Gattiker, Higdon, Keller-McNulty, McKay, Moore,
  and Williams]{gattiker2006combining}
Jim Gattiker, Dave Higdon, Sallie Keller-McNulty, Michael McKay, Leslie Moore,
  and Brian Williams.
\newblock Combining experimental data and computer simulations, with an
  application to flyer plate experiments.
\newblock \emph{Bayesian Analysis}, 1\penalty0 (4):\penalty0 765--792, 2006.

\bibitem[Gelfand et~al.(2004{\natexlab{a}})Gelfand, Schmidt, Banerjee, and
  Sirmans]{gelfand_nonstationary_2004}
Alan~E. Gelfand, Alexandra~M. Schmidt, Sudipto Banerjee, and C.~F. Sirmans.
\newblock Nonstationary multivariate process modeling through spatially varying
  coregionalization.
\newblock \emph{Test}, 13\penalty0 (2):\penalty0 263--312, December
  2004{\natexlab{a}}.

\bibitem[Gelfand et~al.(2004{\natexlab{b}})Gelfand, Schmidt, Banerjee, and
  Sirmans]{gelfand2004nonstationary}
Alan~E Gelfand, Alexandra~M Schmidt, Sudipto Banerjee, and CF~Sirmans.
\newblock Nonstationary multivariate process modeling through spatially varying
  coregionalization.
\newblock \emph{Test}, 13\penalty0 (2):\penalty0 263--312, 2004{\natexlab{b}}.

\bibitem[Genton and Kleiber(2015)]{genton2015cross}
Marc~G Genton and William Kleiber.
\newblock Cross-covariance functions for multivariate geostatistics.
\newblock \emph{Statistical Science}, 30\penalty0 (2):\penalty0 147--163, 2015.

\bibitem[Gibbs and MacKay(1997)]{gibbs_efficient_1997}
Mark Gibbs and David J.~C. MacKay.
\newblock Efficient {Implementation} of {Gaussian} {Processes}.
\newblock Technical report, Cavendish Laboratory, Cambridge, 1997.

\bibitem[Goulard and Voltz(1992)]{goulard1992linear}
Michel Goulard and Marc Voltz.
\newblock Linear coregionalization model: tools for estimation and choice of
  cross-variogram matrix.
\newblock \emph{Mathematical Geology}, 24\penalty0 (3):\penalty0 269--286,
  1992.

\bibitem[Gramacy and Apley(2015)]{gramacy_local_2015}
Robert~B. Gramacy and Daniel~W. Apley.
\newblock Local {Gaussian} {Process} {Approximation} for {Large} {Computer}
  {Experiments}.
\newblock \emph{Journal of Computational and Graphical Statistics}, 24\penalty0
  (2):\penalty0 561--578, April 2015.

\bibitem[Gramacy and Lee(2008)]{gramacy_bayesian_2008}
Robert~B Gramacy and Herbert K.~H Lee.
\newblock Bayesian {Treed} {Gaussian} {Process} {Models} {With} an
  {Application} to {Computer} {Modeling}.
\newblock \emph{Journal of the American Statistical Association}, 103\penalty0
  (483):\penalty0 1119--1130, September 2008.

\bibitem[Gramacy and Lee(2009)]{gramacy_adaptive_2009}
Robert~B. Gramacy and Herbert K.~H. Lee.
\newblock Adaptive {Design} and {Analysis} of {Supercomputer} {Experiments}.
\newblock \emph{Technometrics}, 51\penalty0 (2):\penalty0 130--145, May 2009.

\bibitem[Gramacy and Lee(2012)]{gramacy_cases_2012}
Robert~B. Gramacy and Herbert K.~H. Lee.
\newblock Cases for the nugget in modeling computer experiments.
\newblock \emph{Statistics and Computing}, 22\penalty0 (3):\penalty0 713--722,
  May 2012.

\bibitem[Gramacy and Polson(2011)]{gramacy_particle_2011}
Robert~B. Gramacy and Nicholas~G. Polson.
\newblock Particle {Learning} of {Gaussian} {Process} {Models} for {Sequential}
  {Design} and {Optimization}.
\newblock \emph{Journal of Computational and Graphical Statistics}, 20\penalty0
  (1):\penalty0 102--118, January 2011.

\bibitem[Higdon(2002)]{higdon_space_2002}
Dave Higdon.
\newblock Space and space-time modeling using process convolutions.
\newblock In \emph{Quantitative methods for current environmental issues},
  pages 37--56. Springer, 2002.

\bibitem[Jolliffe(2002)]{jolliffe_principal_2002}
I.~Jolliffe.
\newblock \emph{Principal {Component} {Analysis}}.
\newblock Springer New York, 2 edition, 2002.

\bibitem[Kennedy and O'Hagan(2001)]{kennedy_bayesian_2001}
Marc~C. Kennedy and Anthony O'Hagan.
\newblock Bayesian calibration of computer models.
\newblock \emph{Journal of the Royal Statistical Society: Series B (Statistical
  Methodology)}, 63\penalty0 (3):\penalty0 425--464, January 2001.

\bibitem[Kersting et~al.(2007)Kersting, Plagemann, Pfaff, and
  Burgard]{kersting_most_2007}
Kristian Kersting, Christian Plagemann, Patrick Pfaff, and Wolfram Burgard.
\newblock Most {Likely} {Heteroscedastic} {Gaussian} {Process} {Regression}.
\newblock In \emph{Proceedings of the 24th {International} {Conference} on
  {Machine} {Learning}}, {ICML} '07, pages 393--400, New York, NY, USA, 2007.
  Association for Computing Machinery.

\bibitem[Kim et~al.(2005)Kim, Mallick, and Holmes]{kim_analyzing_2005}
Hyoung-Moon Kim, Bani~K. Mallick, and C.~C. Holmes.
\newblock Analyzing {Nonstationary} {Spatial} {Data} {Using} {Piecewise}
  {Gaussian} {Processes}.
\newblock \emph{Journal of the American Statistical Association}, 100\penalty0
  (470):\penalty0 653--668, June 2005.

\bibitem[Kůrková(1992)]{kurkova_kolmogorovs_1992}
Věra Kůrková.
\newblock Kolmogorov's theorem and multilayer neural networks.
\newblock \emph{Neural Networks}, 5\penalty0 (3):\penalty0 501--506, January
  1992.

\bibitem[L{\'a}zaro-Gredilla and Titsias(2011)]{lazaro2011variational}
Miguel L{\'a}zaro-Gredilla and Michalis~K Titsias.
\newblock Variational heteroscedastic gaussian process regression.
\newblock In \emph{ICML}, 2011.

\bibitem[L{\'a}zaro-Gredilla et~al.(2010)L{\'a}zaro-Gredilla,
  Quinonero-Candela, Rasmussen, and
  Figueiras-Vidal]{lazaro_gredilla_sparse_2011}
Miguel L{\'a}zaro-Gredilla, Joaquin Quinonero-Candela, Carl~Edward Rasmussen,
  and An{\'\i}bal~R Figueiras-Vidal.
\newblock Sparse spectrum gaussian process regression.
\newblock \emph{The Journal of Machine Learning Research}, 11:\penalty0
  1865--1881, 2010.

\bibitem[Le~Méhauté et~al.(1997)Le~Méhauté, Rabut, and
  Schumaker]{le_mehaute_surface_1997}
Alain Le~Méhauté, Christophe Rabut, and Larry~L. Schumaker, editors.
\newblock \emph{Surface fitting and multiresolution methods}.
\newblock Vanderbilt University Press, Nashville, TN, 1st ed edition, 1997.

\bibitem[Lynn and McCulloch(2000)]{lynn_using_2000}
Henry~S. Lynn and Charles~E. McCulloch.
\newblock Using {Principal} {Component} {Analysis} and {Correspondence}
  {Analysis} for {Estimation} in {Latent} {Variable} {Models}.
\newblock \emph{Journal of the American Statistical Association}, 95\penalty0
  (450):\penalty0 561--572, 2000.

\bibitem[Mardia and Goodall(1993)]{mardia_spatial-temporal_1993}
Kanti~V Mardia and Colin~R Goodall.
\newblock Spatial-temporal analysis of multivariate environmental monitoring
  data.
\newblock \emph{Multivariate environmental statistics}, 6\penalty0
  (76):\penalty0 347--385, 1993.

\bibitem[Montúfar et~al.(2014)Montúfar, Pascanu, Cho, and
  Bengio]{montufar_number_2014}
Guido Montúfar, Razvan Pascanu, Kyunghyun Cho, and Yoshua Bengio.
\newblock On the {Number} of {Linear} {Regions} of {Deep} {Neural} {Networks}.
\newblock \emph{arXiv e-prints}, page arXiv:1402.1869, February 2014.

\bibitem[Moré and Wild(2009)]{more_benchmarking_2009}
Jorge~J Moré and Stefan~M Wild.
\newblock Benchmarking derivative-free optimization algorithms.
\newblock \emph{SIAM Journal on Optimization}, 20\penalty0 (1):\penalty0
  172--191, 2009.

\bibitem[Myers(1984)]{myers_co-kriging_1984}
Donald~E. Myers.
\newblock Co-{Kriging} — {New} {Developments}.
\newblock In Georges Verly, Michel David, Andre~G. Journel, and Alain Marechal,
  editors, \emph{Geostatistics for {Natural} {Resources} {Characterization}:
  {Part} 1}, pages 295--305. Springer Netherlands, Dordrecht, 1984.

\bibitem[Nareklishvili et~al.(2022)Nareklishvili, Polson, and
  Sokolov]{nareklishvili2022deep}
Maria Nareklishvili, Nicholas Polson, and Vadim Sokolov.
\newblock Deep partial least squares for iv regression.
\newblock \emph{arXiv preprint arXiv:2207.02612}, 2022.

\bibitem[Neal(1996)]{neal_bayesian_1996}
Radford~M. Neal.
\newblock \emph{Bayesian {Learning} for {Neural} {Networks}}, volume 118 of
  \emph{Lecture {Notes} in {Statistics}}.
\newblock Springer New York, New York, NY, 1996.

\bibitem[Neal(2003)]{neal2003slice}
Radford~M Neal.
\newblock Slice sampling.
\newblock \emph{The annals of statistics}, 31\penalty0 (3):\penalty0 705--767,
  2003.

\bibitem[Osborne et~al.(2009)Osborne, Garnett, and
  Roberts]{osborne2009gaussian}
Michael~A Osborne, Roman Garnett, and Stephen~J Roberts.
\newblock Gaussian processes for global optimization.
\newblock In \emph{3rd international conference on learning and intelligent
  optimization (LION3)}, pages 1--15. Citeseer, 2009.

\bibitem[Pascanu et~al.(2013)Pascanu, Montufar, and
  Bengio]{pascanu_number_2013}
Razvan Pascanu, Guido Montufar, and Yoshua Bengio.
\newblock On the number of response regions of deep feed forward networks with
  piece-wise linear activations.
\newblock \emph{arXiv:1312.6098 [cs]}, December 2013.

\bibitem[Plumlee and Tuo(2014)]{plumlee2014building}
Matthew Plumlee and Rui Tuo.
\newblock Building accurate emulators for stochastic simulations via quantile
  kriging.
\newblock \emph{Technometrics}, 56\penalty0 (4):\penalty0 466--473, 2014.

\bibitem[Polson et~al.(2021)Polson, Sokolov, and Xu]{polson2021deep}
Nicholas Polson, Vadim Sokolov, and Jianeng Xu.
\newblock Deep learning partial least squares.
\newblock \emph{arXiv preprint arXiv:2106.14085}, 2021.

\bibitem[Rasmussen and Williams(2006)]{rasmussen_gaussian_2006}
Carl~Edward Rasmussen and Christopher K.~I. Williams.
\newblock \emph{Gaussian processes for machine learning}.
\newblock Adaptive computation and machine learning. MIT Press, Cambridge,
  Mass, 2006.

\bibitem[Reich et~al.(2011)Reich, Eidsvik, Guindani, Nail, and
  Schmidt]{reich_class_2011}
Brian~J. Reich, Jo~Eidsvik, Michele Guindani, Amy~J. Nail, and Alexandra~M.
  Schmidt.
\newblock A class of covariate-dependent spatiotemporal covariance functions
  for the analysis of daily ozone concentration.
\newblock \emph{Annals of Applied Statistics}, 5\penalty0 (4):\penalty0
  2425--2447, 2011.

\bibitem[Romero et~al.(2013)Romero, Krause, and Arnold]{romero_navigating_2013}
Philip~A. Romero, Andreas Krause, and Frances~H. Arnold.
\newblock Navigating the protein fitness landscape with {Gaussian} processes.
\newblock \emph{Proceedings of the National Academy of Sciences}, 110\penalty0
  (3):\penalty0 E193--E201, 2013.

\bibitem[Sacks et~al.(1989)Sacks, Welch, Mitchell, and Wynn]{sacks_design_1989}
Jerome Sacks, William~J. Welch, Toby~J. Mitchell, and Henry~P Wynn.
\newblock Design and {Analysis} of {Computer} {Experiments}.
\newblock \emph{Statistical Science}, 4\penalty0 (4):\penalty0 409--423,
  November 1989.

\bibitem[Schmidt et~al.(2011)Schmidt, Guttorp, and
  O'Hagan]{schmidt_considering_2011}
Alexandra~M. Schmidt, Peter Guttorp, and Anthony O'Hagan.
\newblock Considering covariates in the covariance structure of spatial
  processes.
\newblock \emph{Environmetrics}, 22\penalty0 (4):\penalty0 487--500, June 2011.

\bibitem[Schultz and Sokolov(2018)]{schultz_bayesian_2018}
Laura Schultz and Vadim Sokolov.
\newblock Bayesian {Optimization} for {Transportation} {Simulators}.
\newblock \emph{Procedia Computer Science}, 130:\penalty0 973--978, 2018.

\bibitem[Schultz et~al.(2022)Schultz, Auld, and Sokolov]{schultz2022bayesian}
Laura Schultz, Joshua Auld, and Vadim Sokolov.
\newblock Bayesian calibration for activity based models.
\newblock \emph{arXiv preprint arXiv:2203.04414}, 2022.

\bibitem[Shan and Wang(2010)]{shan_survey_2010}
Songqing Shan and G.~Gary Wang.
\newblock Survey of modeling and optimization strategies to solve
  high-dimensional design problems with computationally-expensive black-box
  functions.
\newblock \emph{Structural and Multidisciplinary Optimization}, 41\penalty0
  (2):\penalty0 219--241, March 2010.

\bibitem[Snoek et~al.(2014)Snoek, Swersky, Zemel, and Adams]{snoek_input_2014}
Jasper Snoek, Kevin Swersky, Richard~S. Zemel, and Ryan~P. Adams.
\newblock Input {Warping} for {Bayesian} {Optimization} of {Non}-{Stationary}
  {Functions}.
\newblock In \emph{{ICML}}, pages 1674--1682, 2014.

\bibitem[Teh et~al.(2005)Teh, Seeger, and Jordan]{teh_semiparametric_2005}
Yee~Whye Teh, Matthias Seeger, and Michael~I Jordan.
\newblock Semiparametric latent factor models.
\newblock In \emph{International Workshop on Artificial Intelligence and
  Statistics}, pages 333--340. PMLR, 2005.

\bibitem[Viboud et~al.(2018)Viboud, Sun, Gaffey, Ajelli, Fumanelli, Merler,
  Zhang, Chowell, Simonsen, and Vespignani]{viboud_rapidd_2018}
Cécile Viboud, Kaiyuan Sun, Robert Gaffey, Marco Ajelli, Laura Fumanelli,
  Stefano Merler, Qian Zhang, Gerardo Chowell, Lone Simonsen, and Alessandro
  Vespignani.
\newblock The {RAPIDD} ebola forecasting challenge: {Synthesis} and lessons
  learnt.
\newblock \emph{Epidemics}, 22:\penalty0 13 -- 21, 2018.

\bibitem[Wikle(2019)]{wikle2019comparison}
Christopher~K Wikle.
\newblock Comparison of deep neural networks and deep hierarchical models for
  spatio-temporal data.
\newblock \emph{Journal of Agricultural, Biological and Environmental
  Statistics}, 24\penalty0 (2):\penalty0 175--203, 2019.

\bibitem[Wikle and Zammit-Mangion(2022)]{wikle2022statistical}
Christopher~K Wikle and Andrew Zammit-Mangion.
\newblock Statistical deep learning for spatial and spatio-temporal data.
\newblock \emph{arXiv preprint arXiv:2206.02218}, 2022.

\bibitem[Wilson and Adams(2013)]{wilson2013gaussian}
Andrew Wilson and Ryan Adams.
\newblock Gaussian process kernels for pattern discovery and extrapolation.
\newblock In \emph{International conference on machine learning}, pages
  1067--1075. PMLR, 2013.

\bibitem[Wilson et~al.(2014)Wilson, Gilboa, Nehorai, and
  Cunningham]{wilson_fast_2014}
Andrew~G Wilson, Elad Gilboa, Arye Nehorai, and John~P Cunningham.
\newblock Fast kernel learning for multidimensional pattern extrapolation.
\newblock \emph{Advances in neural information processing systems}, 27, 2014.

\bibitem[Yang et~al.(2015)Yang, Wikle, Holan, Myers, and
  Sudduth]{yang_bayesian_2015}
Wen-Hsi Yang, Christopher~K. Wikle, Scott~H. Holan, D.~Brenton Myers, and
  Kenneth~A. Sudduth.
\newblock {Bayesian Analysis Of Spatially-Dependent Functional Responses With
  Spatially-Dependent Multi-Dimensional Functional Predictors}.
\newblock \emph{Statistica Sinica}, 25\penalty0 (1):\penalty0 205--223, 2015.

\bibitem[Álvarez and Lawrence(2011)]{alvarez_computationally_2011}
Mauricio~A. Álvarez and Neil~D. Lawrence.
\newblock Computationally efficient convolved multiple output {Gaussian}
  processes.
\newblock \emph{Journal of Machine Learning Research}, 12\penalty0
  (May):\penalty0 1459--1500, 2011.

\end{thebibliography}
\end{document}